\documentclass[prl, aps, showpacs, groupedaddress, superscriptaddress, twocolumn, numerical]{revtex4-1}

\usepackage[utf8]{inputenc}
\usepackage{color}
\usepackage{bbm} 
\usepackage{nicefrac}
\usepackage{amsfonts,amsmath,amssymb,stmaryrd}
\usepackage{braket}

\usepackage[autostyle=true]{csquotes}

\usepackage{tabularx}
\usepackage{multirow} 
\usepackage{hhline}
\usepackage{graphicx}
\usepackage{subfigure}  %
\usepackage{bbm} 
\usepackage{hyperref}
\usepackage{xcolor}

\usepackage{mathrsfs}
\usepackage{verbatim}
\usepackage{centernot}
\usepackage{ulem}
\usepackage{array}
\usepackage{cancel}
\usepackage{ifthen}
\usepackage{bm}	%
\usepackage{todonotes}
\usepackage{siunitx}
\usepackage{float}
\presetkeys{todonotes}{size=\tiny}{}

\InputIfFileExists{gitinfo-latex.inc}{}{}

\usepackage[acronym,nomain]{glossaries}

\usepackage{units}
\usepackage{upgreek}

\definecolor{azure}{rgb}{.4,.7,.7}

\newcommand{\se}{{SE }}

\newcommand{\hint}{\hat{H}^{\mathrm{int}}}

\newcommand{\mFimp}{m_{F, \mathrm{i}}}
\newcommand{\mFbath}{m_{F, \mathrm{b}}}
\newcommand{\Fimp}{F_{\mathrm{i}}}
\newcommand{\Fbath}{F_{\mathrm{b}}}
\newcommand{\eigenimp}{\ket{\Fimp, \mFimp}}
\newcommand{\eigenbath}{\ket{\Fbath, \mFbath}}
\newcommand{\mtot}{M}

\newcommand{\Eexo}{E_{\mathrm{exo}}}

\newcommand{\spinrate}{\Gamma_{\mathrm{se}}}
\newcommand{\lossrate}{\Lambda}
\newcommand{\Gammael}{\Gamma_{\mathrm{el}}}
\newcommand{\Gel}{{G_\mathrm{el}}}
\newcommand{\Ttwoprime}{T_2^{\prime}}
\newcommand{\nbath}{n_{\mathrm{BEC}}}
\newcommand{\nimp}{n_{\mathrm{i}}}
\newcommand{\nexp}{\left<n\right>}

\definecolor{color}{RGB}{0,0, 0} %
\renewcommand{\sout}[1]{\unskip}

\newcommand{\tukl}{Department of Physics and Research Center OPTIMAS, Technische Universit\"at Kaiserslautern, Germany}

\makeatletter
\newsavebox\myboxA
\newsavebox\myboxB
\newlength\mylenA
\newcommand*\xoverline[2][0.75]{%
	\sbox{\myboxA}{$\m@th#2$}%
	\setbox\myboxB\null%
	\ht\myboxB=\ht\myboxA%
	\dp\myboxB=\dp\myboxA%
	\wd\myboxB=#1\wd\myboxA%
	\sbox\myboxB{$\m@th\overline{\copy\myboxB}$}%
	\setlength\mylenA{\the\wd\myboxA}%
	\addtolength\mylenA{-\the\wd\myboxB}%
	\ifdim\wd\myboxB<\wd\myboxA%
	\rlap{\hskip 0.5\mylenA\usebox\myboxB}{\usebox\myboxA}%
	\else
	\hskip -0.5\mylenA\rlap{\usebox\myboxA}{\hskip 0.5\mylenA\usebox\myboxB}%
	\fi}
\makeatother

\begin{document}
	\title{Quantum spin dynamics of individual neutral impurities \\ coupled to a Bose-Einstein condensate}
	
	\author{Felix Schmidt}
	\affiliation{\tukl}

	\author{Daniel Mayer}
	\affiliation{\tukl}
	
	\author{Quentin Bouton}
	\affiliation{\tukl}
	
	\author{Daniel Adam}
	\affiliation{\tukl}

	\author{Tobias Lausch}
	\affiliation{\tukl}
	
	\author{Nicolas Spethmann}
	\email{present address: Physikalisch-Technische Bundesanstalt, Bundesallee 100, 38116 Braunschweig, Germany}
	\affiliation{\tukl}
	
	\author{Artur Widera}
	\email{email: widera@physik.uni-kl.de}
	\affiliation{\tukl}
	\affiliation{Graduate School Materials Science in Mainz, Gottlieb-Daimler-Strasse 47, 67663 Kaiserslautern, Germany}

	\date{\today}
	
	\begin{abstract}
	We report on spin dynamics of individual, localized neutral impurities immersed in a Bose-Einstein condensate. Single Cesium atoms are transported into a cloud of Rubidium atoms, thermalize with the bath, and the ensuing spin-exchange between localized impurities with quasi-spin $\Fimp=3$ and bath atoms with $\Fbath=1$ is resolved. Comparing our data to numerical simulations of spin dynamics we find that, for gas densities in the BEC regime, the dynamics is dominated by the condensed fraction of the cloud. We spatially resolve the density overlap of impurities and gas by the spin-population of impurities.
	Finally we trace the coherence of impurities prepared in a coherent superposition of internal states when coupled to a gas of different densities. For our choice of states we show that, despite high bath densities and thus fast thermalization rates, the impurity  coherence is not affected by the bath, realizing a regime of sympathetic cooling while maintaining internal state coherence.
	Our work paves the way toward non-destructive probing of quantum many-body systems via localized impurities.
\end{abstract}
\maketitle
Individual impurities interacting with a quantum system form a paradigmatic model system of quantum physics, with numerous applications in probing, quantum state engineering or quantum simulation. Proposals employing the local interaction of atomic impurities in a many-body system include the measurement of various moments \cite{Elliott2016}, excitations \cite{Hangleiter2015} and correlations \cite{Streif2016} of a BEC%
; the cooling of qubits while preserving internal state coherence \cite{Daley2004}; or the study of quasi-particles in novel regimes \cite{Levinsen2017, Grusdt2018}.
Experimentally, ensembles of impurities have been studied in imbalanced quantum gas mixtures for thermometry in the regime of weak coupling \cite{Olf2015}, or for the study of fermionic \cite{Schirotzek2009, Kohstall2012} and bosonic polarons \cite{Jorgensen2016, Hu2016} in the strong coupling limit. Recently, the non-equilibrium spin dynamics of a BEC coupled to a Fermi-gas was investigated interferometrically \cite{Cetina2016} as well as the bath-induced decay of motional coherence of fermions in a trap \cite{Scelle2013}.

Entering the limit of individual impurity atoms facilitates tracing interactions in the single particle limit \cite{Kindermann2016, Hohmann2017}, being sensitive to individual trajectories and rare events \cite{Kindermann2017}. 
Moreover, single-particle control has enabled quantum simulation \cite{Bernien2017, Gross2017} and state engineering \cite{Serwane2011,  Kaufman2015} in a bottom-up approach.
The immersion of individual impurities into a quantum system will transfer this control to many-body physics opening fascinating perspectives for quantum engineering by, e.g., bath-mediated entanglement \cite{Klein2005, Klein2007}. %
Individual impurities have been immersed into BEC as charged particles, either as individual ions \cite{Zipkes2010, Schmid2010}, or as quasi-free charged particles in Rydberg atoms \cite{Balewski2013, Kleinbach2018}. In a lattice approach, the position dynamics of impurities in Bose-Hubbard lattice systems have been studied, where the spin-degree-of-freedom was used to identify the impurity \cite{Fukuhara2013}.
However, the immersion of individual, thermalized impurities with a spin-degree of freedom into a BEC is so far unreported.

Here, we present a hybrid quantum system, comprising a BEC of $^{87}$Rubidium (Rb) atoms and individual neutral impurities of $^{133}$Cesium (Cs).
Independent position control of the impurities is obtained by species-selective optical fields \cite{Schmidt2016_2}. Importantly, the internal impurity state acts as a quasi-spin, which can be coherently manipulated as well as precisely detected.
We study spin dynamics of impurities coupled to ultracold gases at various densities in two different regimes. 

\begin{figure*}
	\begin{center}
		\includegraphics[width=.9\textwidth]{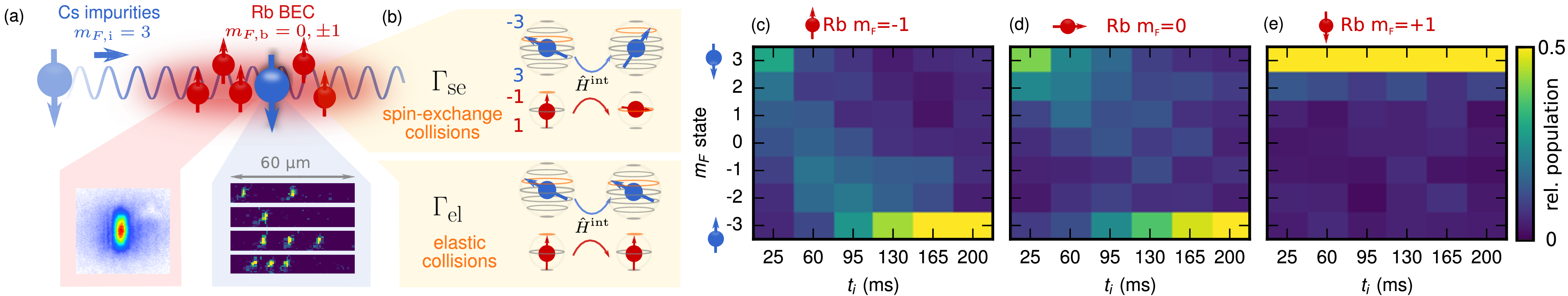}
	\end{center}
	\caption{\textbf{Experiment overview}. (a) Sketch of Cs impurities (blue dots and fluorescence images), immersed in a Rb BEC (red dots and time-of-flight image) by a species-selective optical lattice (blue).
			(b) Sketch of possible interaction paths: elastic (spin-exchange) collisions occur at rate $\Gammael$ ($\spinrate$).
			(c, d, e) Spin evolution of the impurity, prepared in $\mFimp=3$ in a thermal Rb bath in states $\mFbath=-1$, $0$, $1$ (populations normalized column wise) \textcolor{color}{with density overlap $\nexp=\SI{2.6e12}{\centi \meter^{-3}}$ and magnetic field of $\SI{250}{\milli G}$}.
			The Zeeman energy of Cs and Rb determines the direction of spin-exchange (SE).
			For $\mFbath = 0, -1$ (c,d) \se between impurity and bath atoms lowers $\mFimp$, so impurity atoms are eventually pumped to $\mFimp = -3$.
			For Rb in $\mFbath=1$ (e), this process is energetically forbidden.
			\textcolor{color}{For $\mFbath=-1$ (c), we expect SE at a time scale of $1 /\spinrate = 1 / (G \nexp) = \SI{22}{\milli \second}$, leading to SE pumping of Cs from $\mFimp=3$ to $\mFimp=-3$ within $6 / \spinrate \approx \SI{130}{\milli \second}$.}
			}
	\label{fig:Fig1Overview}
\end{figure*}

First, we study spin-exchange \textcolor{color}{(SE)} dynamics of individual, localized impurities immersed in the gas, forming the bosonic analogue to \se in a fermionic gas \cite{Riegger2018}. 
\se interaction changes the impurities' spin-populations, rendering the impurity spin an effective memory for the number of elastic collisions. 
Thereby, we demonstrate successful immersion of individual impurities into a BEC from directly measuring $s$-wave collisions rather than three-body recombination. 
Second, we prepare the impurities in a coherent superposition of internal states. While bath-mediated decoherence has been observed previously \cite{Ratschbacher2013, Scelle2013}, we here realize a regime where the motion of a qubit can be efficiently cooled sympathetically by frequent collision with the gas, while internal-state coherence is preserved.
Furthermore, our ability to detect both, the kinetic energy distribution \cite{Alt2003, Hohmann2016} as well as local, internal-state coherence of an impurity in a bulk BEC will open the door to study relaxation of non-equilibrium quantum systems  \cite{Eisert2015}, and to compare and control the different time-scales associated with spin- or motional relaxation for quantum state engineering.

For low magnetic fields, the state of an impurity atom $i$ (bath atom $b$) is given by the hyperfine state $\eigenimp$ ($\eigenbath$), with total angular momentum quantum number $\Fimp=3, 4$ ($\Fbath=1$) and the projection onto the quantization axis $\mFimp$ ($\mFbath$).
The interaction of impurity and bath atoms is given by the central molecular interaction potential. Hyperfine interaction couples collisional channels, defined by the total angular momentum $\mathbf{F} = \mathbf{F}_{\mathrm{i}} + \mathbf{F}_{\mathrm{b}}$ and its projection $M$. 
The interaction Hamiltonian for low-energy $s$-wave collisions writes \cite{Ho1998} 
\begin{equation}
\hat{H}^{\mathrm{int}} = \left< n\right> \sum_{F=\left| \Fimp - \Fbath \right|}^{\Fimp + \Fbath} g_F \mathcal{P}_F .
\label{eq:interactionHamiltonian}
\end{equation}
Here, $\mathcal{P}_F = \sum_{M=-F}^{F} \ket{\Fimp \Fbath; F, M}\bra{\Fimp \Fbath; F, M}$ are the projection operators onto the total spin $F$, $\nexp$ is the spatial wave function overlap, and
$g_F = \frac{4 \pi \hbar^2 }{\mu} a_F$ is the coupling constant with $s$-wave scattering length $a_F$ in the scattering channel $F$ and reduced mass $\mu$ (for details see appendix\nocite{Lausch2018, Kempen2002, Takekoshi2012, Davis1995, Hohmann2017, Stoof1988, Tiemann2018, Treutlein2001, Dalfovo1999, Kagan1996, Castin1996, Kolovsky2004, LeBlanc2007, Schmidt2016, Kuhr2001, Kuhr2005}).

Hamiltonian (\ref{eq:interactionHamiltonian}) allows for three distinct processes \cite{Stoof1988}: 
First, elastic collisions preserve the internal states and lead to a fast thermalization of the impurity, induced by the first collision with a bath atom \cite{Hohmann2017}.
For elastic interaction, eq.~(\ref{eq:interactionHamiltonian}) sums the contributions of the involved scattering channels with total spin $F$, and the interaction energy $E_{\mathrm{el}} = 4 \pi \hbar^2 a / \mu \cdot  \nexp$ is quantified by the effective scattering length $a=648\, a_0$ with the Bohr radius $a_0$ appendix.

Second, eq.~(\ref{eq:interactionHamiltonian}) couples collisional channels of different spin-states and the respective energy scale is given by a weighted difference in respective $F$ channels.
Such \se processes maintain the total projection $\mtot = \mFimp + \mFbath$ and can lead to a spin transfer from an impurity to a bath atom in quanta of $\pm \hbar, \pm 2 \hbar$ (with reduced Planck's constant $\hbar$) for our system (see fig.~\ref{fig:Fig1Overview}(b)-(e)).
In a \se collision between individual impurities and the BEC, a single bath atom is projected from the spinor mode, initially populated by the BEC, into an empty spinor (vacuum) mode with different $\mFbath$. 
The excitation of the final state in a \se process is hence governed by single-particle rather than Bogoliubov dispersion.
Nevertheless, beyond collective excitations, the strongly modified density of the BEC compared to a thermal gas provides a clear signature of the BEC that is revealed in the \se rate.
\textcolor{color}{\se strongly depends on the internal Rb state (see fig.~\ref{fig:Fig1Overview}(c-e)), which can be utilized either to probe the Rb density by explicitly allowing \se for $\mFbath=-1$, or to disable SE for $\mFbath=1$, both applied below.}

\textcolor{color}{\sout{Finally }Third}, for impurities prepared \textcolor{color}{\sout{in a superposition state} in a superposition of internal states, here $\ket{g}=\ket{\Fimp=3, \mFimp=3}$ and $\ket{e}=\ket{\Fimp=4, \mFimp=3}$,} the difference of the elastic interaction energies for the two internal states leads to a relative phase shift between both states with an effective scattering length difference $\Delta a_{\mathrm{g-e}} = -33 \, a_0$ \textcolor{color}{\sout{for the qubit states discussed later}}
\textcolor{color}{, which leads to bath-induced inhomogeneous dephasing of the impurity qubit due to the inhomogeneous bath density distribution} (see appendix).

The interaction strengths are quantified by the microscopic rate constants $\Gel$ ($G$) for elastic (SE) collisions, which depend on the internal states of both collision partners, where the assumption of state-independent rate constants is applicable within a certain range, given in our experiment (see appendix).
Experimental observables are the rates of elastic collisions $\Gammael=\Gel\nexp$ and \se $\spinrate=G \nexp$, which are calculated from the density overlap $\nexp  = \int |\phi_{\mathrm{i}}|^2 n_{\mathrm{b}} \mathrm{d}^3 r$ of both species.
Here,  $\phi_{\mathrm{i}}$ is the impurity wave function and $n_{\mathrm{b}} = |\phi_{\mathrm{b}}|^2 + n_{\mathrm{b, therm}}$ the Rb bath density, including the condensate wave function $\phi_{\mathrm{b}}$ and the thermal BEC background $n_{\mathrm{b, therm}}$.
Typically, the rates of elastic collisions and \se have a fixed ratio $\nicefrac{\Gammael}{\spinrate} = \nicefrac{\Gel}{G} \approx 11$, which allows to infer the number of elastic collisions from the number of \se events (see appendix).

Experimentally (see Fig.~\ref{fig:Fig1Overview}~(a)), we prepare individual neutral Cs impurities inside a Rb BEC by first creating a BEC of about $10^4$ Rb atoms at $\SI{300}{\nano K}$ with a typical condensate fraction of about $0.3$ in a crossed dipole trap (see appendix).
Single or up to ten Cs atoms are cooled and trapped in a high-gradient magneto-optical trap. 
Cs is transferred into a second crossed dipole trap, sharing the horizontal (along $z$ axis) dipole trap beam with Rb.
Using microwave radiation, we prepare Rb in the $\Fbath=1$ state with $ \mFbath=0, \pm 1$.
Cs impurities are further cooled and compressed by degenerate Raman sideband cooling \cite{Treutlein2001}, leaving Cs at a temperature of approximately $\SI{2}{\micro \kelvin}$ in the absolute ground state $\ket{\Fimp=3, \mFimp=3}$.
The interaction of impurities and BEC is initiated by transporting Cs atoms into the BEC  with a one-dimensional, species-selective conveyor belt lattice, achieving position control over impurity atoms independently from the Rb bath \cite{Schmidt2016_2}.
In the lattice, impurities are localized within the BEC along the main trap axis $z$, allowing spatial resolution in the experiment.
Inelastic three-body recombination (Cs-Rb-Rb) can lead to Cs loss.
By turning off the axial BEC confinement, we induce an expansion of the BEC $\SI{6.5}{\milli \second}$ prior to the interaction. 
This reduces the BEC density (see Fig.~\ref{fig:Fig2BECThermal}~(b)) and thereby enhances the impurity lifetime (see appendix).
After a given interaction duration $t_i$, Rb is removed from the trap by means of a resonant light pulse.
The impurity position is measured in the lattice by fluorescence imaging and the spin population $\mFimp$ is read out \textit{in situ} (see appendix and Fig.~\ref{fig:Fig1Overview}~(c)-(e)).

\begin{figure}[tb]
	\begin{center}
		\includegraphics[width=70mm]{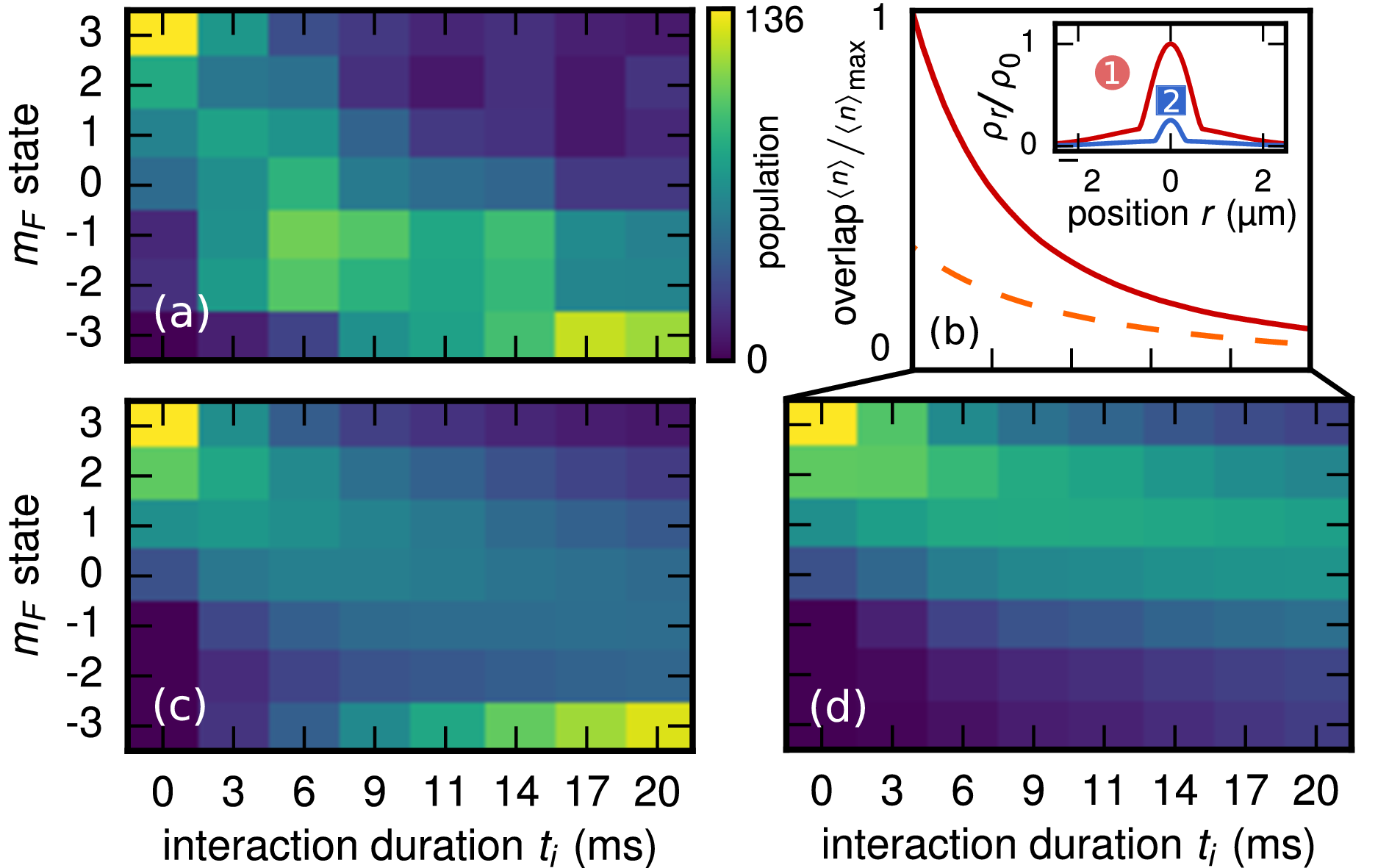}
	\end{center}
	\caption{(a) Measured \se dynamics of Cs impurities (initially in $\mFimp=3$) in an expanding BEC ($\mFbath=-1$)
	for a magnetic background field of $\SI{250}{\milli G}$.
			Initial population of Cs atoms in $\mFimp=3$ is 136.
			(b) Calculated averaged density overlap $\nexp$ of impurities with condensate fraction (red) and  thermal background (orange, dashed) of the BEC shows  enhanced interaction with the condensate fraction ($\nexp_{\mathrm{max}} = \SI{2.7e13}{\centi \meter^{-3}}$).
			$\nexp$ decreases due to the BEC expansion.
			The inset shows the radial BEC line density $\rho_r$ ($\rho_0=\SI{440}{\micro \meter^{-1}}$) at $t_i = 0$ (marked as 1) and $t_i=\SI{20}{\milli \second}$ (2).
			From this, we calculate the spin evolution of Cs in the bath (no free parameters; $G=\SI{1.71e-11}{cm^3 Hz}$) with (c) BEC properties from (a), and (d) a purely thermal Rb cloud at the same temperature and atom number (model details in appendix).
			\textcolor{color}{Data was taken in 3200 experimental runs.}
		}
	\label{fig:Fig2BECThermal}
\end{figure}
We apply our technique to resolve \se dynamics between localized Cs impurities and the Rb bath.
We find that \se successively pumps impurities to the $\mFimp = -3$ state within approximately $\SI{20}{\milli \second}$ (see fig. \ref{fig:Fig2BECThermal}~(a)).
We compare the full measured dynamics to a Monte-Carlo simulation, where the time evolution of all $\mFimp$ populations is modeled assuming fully thermalized impurities within the BEC (see appendix).
The rate constant $G = \SI{1.71(65)e-11}{\hertz \, \centi \meter^3}$ used in the model has been determined in an independent measurement, %
and is consistent with theoretical estimates (see appendix).
Our model (see Fig.~\ref{fig:Fig2BECThermal}~(c)) yields good agreement with the measurement and we attribute deviations to the assumption of a $\mFimp$-independent \se constant $G$ in our model.
From our model we infer that thermalized Cs impurities interacting with a BEC experience the high density at the trap center, enhancing \se with the condensate by a factor of 2.4 with respect to the thermal Rb background.
In fact, a model of the impurities' spin dynamics in a purely thermal Rb bath of same atom number and temperature (shown in Fig.~\ref{fig:Fig2BECThermal}~(d)) cannot explain the measured data due to the lower density overlap.
Thus, the enhanced density overlap of the condensate dominates the interaction even for relatively low condensate fractions in the sample and our data demonstrates the successful immersion of Cs impurities into a \textcolor{color}{\sout{quantum bath} Rb BEC}.
\begin{figure}[b]
	\begin{center}
		\includegraphics[width=85mm]{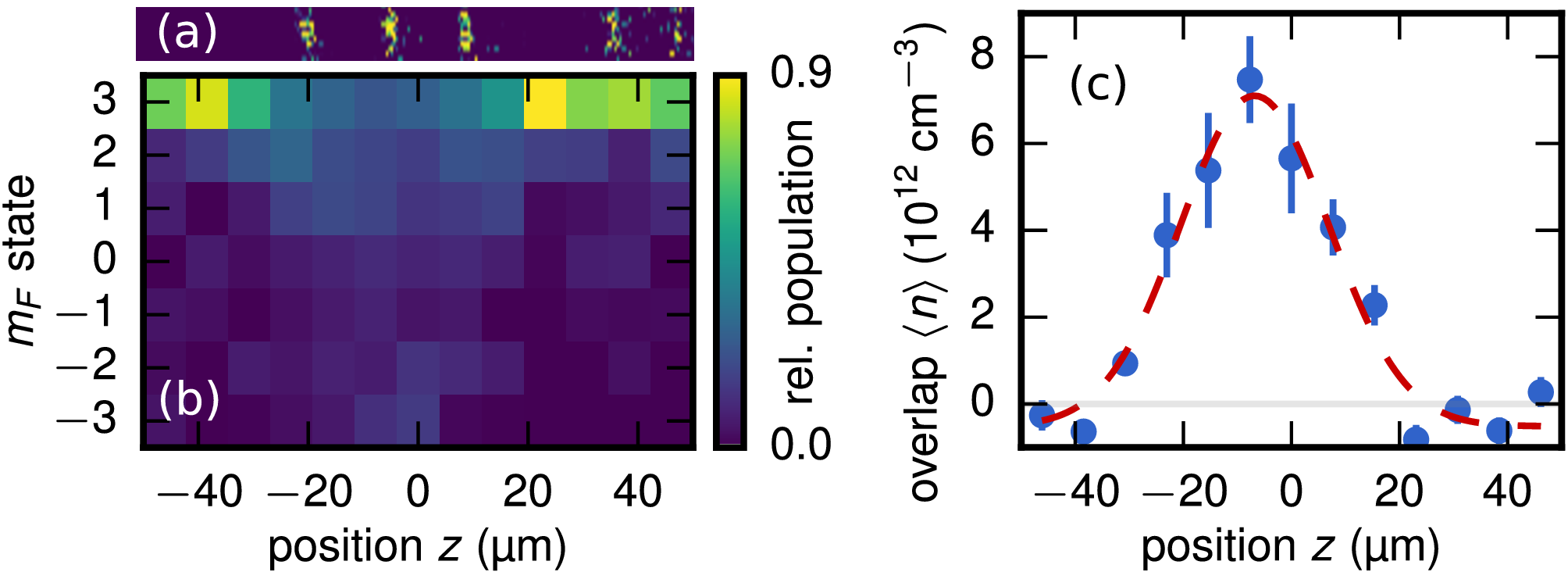}
	\end{center}
	\caption{(a) Position resolved impurity spin dynamics (fluorescence image) in $\mFimp=3$ immersed into a BEC ($\mFbath=-1$) from right to left, without BEC expansion.
			Impurity positions are sorted into bins of $\approx \SI{8}{\micro \meter}$ in order to enhance statistics (optical resolution $\SI{2}{\micro \meter}$). 
		 (b) $\mFimp$ populations are read out after $t_i = \SI{10}{\milli \second}$ (columnwise normalized).
		 Effective density overlap $\left< n \right>$ extracted for each bin from our model (blue dots), using a $\chi^2$ optimization (fitting uncertainty yields error bars).
		The BEC position is extracted in a simple Gaussian fit (red, dashed), where $z=0$ is given by the experimentally determined BEC position via absorption imaging on a different imaging axis, with a possible systematic error of $\approx\SI{10}{\micro \meter}$.
		Impurity transport through the Rb BEC takes $\approx \SI{2}{\milli \second}$ and can lead to \se prior to the beginning of interaction.
	}
	\label{fig:SpatialResolution}
\end{figure}
From our findings, two interesting perspectives emerge for the application of \se.
First, the impurity spin state acts a memory for the number of collisions in the system, allowing to study e.g.~the thermalization of Cs in the bath on a single-particle level \cite{Lausch2018}.
Second, \se at the single-particle level reveals intricate details of the molecular interaction potential with energy scales as low as few $h \cdot \si{\kilo \hertz}$ \cite{Takekoshi2012}.

Next, we employ the localization of Cs impurities in the species-selective lattice to obtain spatially resolved information from the impurities.
We transport individual Cs impurities to selected positions in the vicinity and inside of the Rb cloud, here without relaxing the axial Rb confinement.
Figure \ref{fig:SpatialResolution} shows the position resolved impurity spin population after a fixed interaction duration of $\SI{10}{\milli \second}$, where atomic positions have been binned to $\SI{8}{\micro \meter}$ intervals.
From the $\mFimp$ population in each position bin we extract the impurity-BEC overlap (see fig. \ref{fig:SpatialResolution}~(c)).

\begin{figure}[tb]
	\begin{center}
		\includegraphics[width=.48\textwidth]{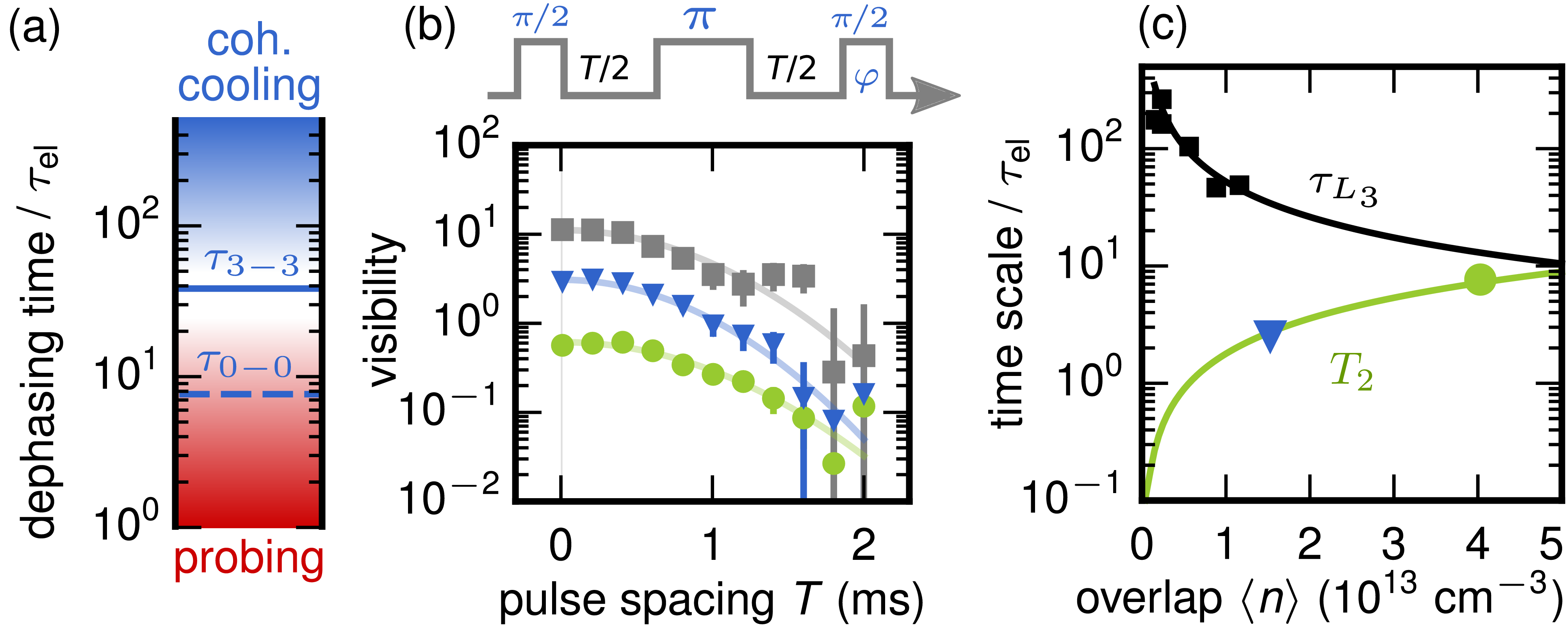}
	\end{center}
	\caption{%
		(a) Expected collisional dephasing time scale of the $\ket{g}=\ket{\Fimp=3,\mFimp=3}$ - $\ket{e}=\ket{4,3}$ qubit ($\tau_{3-3}$), as used for measurement (b), and the $\ket{g}=\ket{3,0}$ - $\ket{e}=\ket{4,0}$ qubit ($\tau_{0-0}$) for comparison, given in multiples of the inverse elastic $s$-wave collision rate ($\tau_{\mathrm{el}}$).
		(b) Pulse-echo sequence applied to Cs impurities, while immersed in Rb bath, which has been expanded for 6 ms prior to impurity immersion.
		Visibility of Ramsey fringes for impurities in a BEC (green circles, \textcolor{color}{$\nexp=\SI{4.0e13}{\centi \meter^{-3}}$}), in a thermal Rb bath (blue triangles, \textcolor{color}{$\nexp=\SI{1.5e13}{\centi \meter^{-3}}$}) and without Rb (gray squares).
		The latter datasets are offset by a constant factor of 5 and 10, respectively.
		\textcolor{color}{Extracted $T_2$ coherence times are given in the text,}
		 error bars give statistical uncertainties.
		\textcolor{color}{(c) Ratio of the experimentally obtained coherence times from from (b) and the elastic collision time $\tau_{\mathrm{el}}$ as a function of the density overlap $\nexp$, showing coherent cooling with $T_2 / \tau_{\mathrm{el}} \gg 1$.
			Green solid line gives the averaged $T_2=\SI{1.1}{\milli \second}$ as a constant.
			Three-body loss (here ratio of three-body loss and elastic collision time) is characterized independently (black data points) and does not limit the coherence measurement.
		\sout{(c) Comparison of , extracted from (b) in various density regimes (markers correspond to (b)).
		The green shaded area is accessible for experiments.		
		Relevant timescales are elastic collisions for thermalization (gray, solid) and atom loss due to three-body recombination (black), where the latter has been determined in an independent measurement.}}
	}
	\label{fig:Fig4Ramsey}
\end{figure}

An important question regards the coherent dynamics of an impurity qubit immersed into the BEC, which is in general determined by dephasing ($T_2$ time) and longitudinal decoherence ($T_1$ time) sources, where the latter is negligible in our system (see ref. \textcolor{color}{\cite{NoteLifetime}} and appendix). 
While bath-mediated decoherence of individual impurities has been studied \cite{Ratschbacher2013, Scelle2013}, here we realize the complementary regime of fast impurity thermalization in the bath, where decoherence is only limited by coupling to external fields despite frequent bath collisions.
Time scales related to impurity-bath interaction (see fig.~\ref{fig:Fig4Ramsey}~(a)) are set by the inverse rate of elastic collisions $\tau_{\mathrm{el}} = 1 / \Gammael = 1 / (\Gel \left<n\right>)$, leading to thermalization of the impurity \cite{Hohmann2017} \textcolor{color}{and the inverse rate of inhomogeneous, bath-induced dephasing $\tau_{\mathrm{g-e}}$, resulting from state-dependent interaction energies.}
For small \textcolor{color}{\sout{bath-induced}} dephasing rates $\tau_{\mathrm{g-e}} \gg \tau_{\mathrm{el}}$ and $T_2 \gg \tau_{\mathrm{el}}$ the qubit can be efficiently cooled while preserving the internal state \cite{Daley2004}.
By contrast, if the associated time scales are of \textcolor{color}{similar} order ($\tau_{\mathrm{g-e}} \geq \tau_{\mathrm{el}}$ and $T_2 \gg \tau_{\mathrm{el}}, \tau_{\mathrm{g-e}}$), the relative phase of the qubit is sensitive to atomic collisions, which can be employed for BEC probing \cite{Hangleiter2015, Elliott2016, Streif2016, Nokkala2016} or bath-mediated impurity entanglement \cite{Klein2005}.

We select the two-level system of hyperfine states $\vert g \rangle \equiv \vert \Fimp=3, \mFimp = 3\rangle$ and $\vert e \rangle \equiv \vert \Fimp=4, \mFimp=3 \rangle$  (see fig. \ref{fig:Fig4Ramsey}~(b)). 
This combination is amenable to differential light shifts, which allows for tight control via state-dependent optical fields, but also causes dephasing \cite{Karski2009_2}. 
We therefore perform a spin-echo sequence to compensate for contributions of quasi-constant dephasing sources.
The visibility $\nu(T)$ at a given free evolution time $T$ is determined by varying the phase $\varphi$ of the second $\pi / 2$-pulse in the sequence.
For Gaussian distributed fluctuations of the transition energy between $\ket{g}$ and $\ket{e}$, we expect a decay of visibility as
$\nu(T) = \nu_{0} \exp{ \left(\nicefrac{-T^2}{T_2^2} \right)}$,
with initial visibility $\nu_0$ \cite{Kuhr2005}.
In figure~\ref{fig:Fig4Ramsey}~(b) we compare coherence measurements in different scenarios.
We immerse ground state impurities into the Rb bath at approximately $\SI{300}{\nano \kelvin}$ to enable thermalization in the bath and perform the spin-echo.
Population loss due to \se is avoided by choosing $\mFbath=1$ for the bath. 
We extract coherence\textcolor{color}{\sout{s}} times in three bath density regimes, see Fig.~\ref{fig:Fig4Ramsey}~\textcolor{color}{\sout{(a)}(b)}, which are $T_2=\SI{1.07\pm 0.10}{\milli \second}$, when Rb is removed before the spin echo, $T_2=\SI{1.07\pm 0.08}{\milli \second}$ in the presence of a thermal bath, and $T_2=\SI{1.17\pm0.06}{\milli \second}$ when immersed into a BEC.
The decoherence is limited by fluctuations of the external magnetic field in the sub-$\si{\milli G}$ range.
Eventually, the qubit population will decay due to three-body recombination at rate $\Gamma_{\mathrm{loss}}$, which is however suppressed by a factor of $2$ \cite{Cheiney2018} in the condensate.
In addition, the contrast $\nu(T)$ is insensitive to the absolute impurity atom number, so we are not fundamentally limited by three-body loss.
The coherence at different densities is compared to the mean-free time between collisions ($\tau_{\mathrm{el}} = \SI{140}{\micro \second}$ for the BEC) as well as the expected bath-induced dephasing (\textcolor{color}{\sout{$\tau_{\mathrm{g-e}} = \SI{6}{\milli \second}$}$\tau_{\mathrm{3-3}} = \SI{6}{\milli \second}$}) in Fig.~\ref{fig:Fig4Ramsey}~(c).
Clearly, our system is in the regime of coherence-maintaining cooling, where $\tau_{\mathrm{3-3}} \gg \tau_{\mathrm{el}}$ and $T_2 \gg \tau_{\mathrm{el}}$, while not affecting the bath. 
In fact, the low-density, thermal bath does barely provide sufficient collisions to feature thermalization \cite{Hohmann2017}, while the higher-density BEC bath ensures thermalization well within the coherence time.
The coherence dynamics is strongly governed by the specific choice of impurity states. For example, preparing the impurity in magnetic-field insensitive \textcolor{color}{\sout{impurity hyperfine states} qubit states $\ket{g}=\ket{3,0}$ - $\ket{e}=\ket{4,0}$} not only extends the coherence time 
(\textcolor{color}{$T \gg \tau_{\mathrm{g-e}}$}) \cite{Kuhr2005}, but also \textcolor{color}{\sout{increases the state-sensitive impurity-bath interaction} enhances bath-induced dephasing} $\delta_{\mathrm{0-0}} = 2\pi / \tau_{0-0}$ by a factor of 10 (see appendix), which is well within the regime \textcolor{color}{\sout{for quantum probing} for probing of BEC properties \cite{Hangleiter2015, Elliott2016, Streif2016, Nokkala2016}} or BEC-mediated entanglement \cite{Klein2005}, \textcolor{color}{see} Fig.~\ref{fig:Fig4Ramsey}(a).

Concluding, we have observed the spin dynamics of individual atoms coupled to an ultracold bath. Tracing the dynamics of spin-exchange we find that the dynamics is dominated by interactions with high density regions of a BEC.
The immersion of individual impurities into a BEC opens the route to couple two BEC states in different $m_F$ states via spin-exchange with a localized impurity. The process can be tuned resonant by microwave dressing, realizing the basic building block for the \textcolor{color}{\sout{B}b}osonic analogue of the Kondo effect. 
Furthermore, using different combination of internal impurity states, the superposition phase can be rendered susceptible to collisions with the bath.
Additionally, relevant scattering lengths and thereby bath-induced interactions are tunable via, e.g., magnetic Feshbach resonances, thereby enabling impurity-aided BEC probing \cite{Hangleiter2015, Elliott2016, Streif2016, Nokkala2016}.
Finally, it will be interesting to study the thermalization dynamics of non-equilibrium or driven quantum systems \cite{Eisert2015} in both motional and spin degrees of freedom, as the local relaxation of the impurity is faster than the global relaxation of the bath due to a strong difference in intra- versus inter-species scattering lengths.
\section*{Acknowledgements}
We thank Michael Hohmann and Farina Kindermann for their help in initially constructing the experiment and for initial discussions, Steve Haupt for his support in preparing the measurements, and Eberhard Tiemann and Axel Pelster for helpful discussions.
A.W. thanks Dieter Meschede for support in initiating the project. 
This work was funded in the early stage by the European Union via ERC Starting grant "QuantumProbe"; equipment was partially contributed by Deutsche Forschungsgemeinschaft via Sonderforschungsbereich (SFB) SFB/TRR185. 
D.M. acknowledges funding  via  SFB/TRR49,  T.L. acknowledges funding by Carl Zeiss Stiftung, and F.S. acknowledges funding by the Studienstiftung des deutschen Volkes.
\bibliography{bibliography_spinexchange}

\begin{thebibliography}{54}
\bibitem [{Not()}]{NoteLifetime}%
  \BibitemOpen
  \href@noop {} {\bibinfo  {journal} {A transition between our qubit hyperfine
  states is dipole-forbidden, thus spontaneous decay is absent on relevant
  experimental time scales}\ }\BibitemShut {NoStop}%
\end{thebibliography}

\section{Appendix}
\subsection{Interaction Hamiltonian and eigenenergies}
\label{sec:interactionPotential}
Our impurity-bath system can be considered an ultracold mixture experiment with extreme imbalance, where the interaction potentials are well known and impurities rapidly thermalize with the Rb bath, while impurity-impurity interaction is negligible.

\begin{figure}
	\centering
	\includegraphics[scale=1]{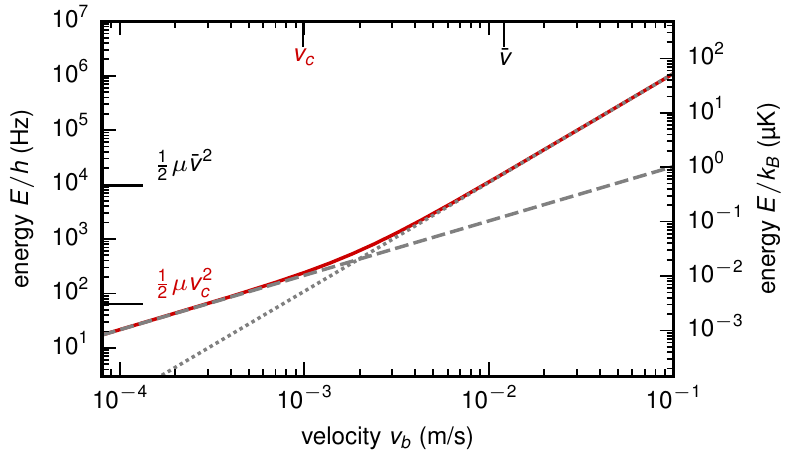}
	\caption{
		Bogoliubov dispersion relation of a Rb BEC for a density of $n = \SI{2.7e13}{\centi \meter^{-3}}$ (red), as used in the measurement of Fig. 2 in the body of this work with $v_b = \hbar k / m_{\mathrm{b}}$.
		The dashed (dotted) line describe the wave-like, linear (particle-like, quadratic) excitation character for low (high) momenta.
		The corresponding speed of sound $v_c$ and the expectation value of the relative collision velocity $\bar{v}$ at $\SI{300}{\nano \kelvin}$ are shown together with the respective energy scales.
		}
	\label{fig:S0}
\end{figure}

Important interaction energy scales in our system are set by the elastic (spin-maintaining) and spin-exchanging collisions between the Cs impurities of mass $m_{\mathrm{i}}$ and bath atoms of mass $m_{\mathrm{b}}$.
The former lead to thermalization of the impurities within the bath, while the latter is a dissipative process, releasing energy into the system.
Thermalization of an impurity in a quantum bath is not trivial, since the impurities' kinetic energy is dissipated by phonon scattering within the BEC \cite{Lausch2018}. 
However, for typical velocity and energy scales in our system, we can assume a particle-like character of elastic s-wave collisions, as discussed in the following.
The dispersion relation $\epsilon_k = \hbar k v_c (1 + 
\nicefrac{\xi^2 k^2}{2})^{1/2}$ of the weakly interacting Rb BEC is shown in fig.~\ref{fig:S0}.
Here, $k = p_b / \hbar$ is the wave vector of Bogoliubov excitations with momentum $p_b$, $v_c = \hbar / (\sqrt{2} \xi m_{\mathrm{b}})$ denotes the critical velocity and $\xi = 1 / \sqrt{8 \pi n a_{\mathrm{bb}}}$ is the BEC healing length at density $n$ with the boson-boson s-wave scattering length $a_{\mathrm{bb}} = 101\,a_0$ and Bohr radius $a_0$ \cite{Kempen2002}.
In a classical bath at temperature $T$, the expectation value for relative collision velocities $\bar{v}$ is given by $\bar{v} = \sqrt{8 k_B T / (\pi \mu)}$ with reduced mass $\mu =\nicefrac{m_{\mathrm{i}} \, m_{\mathrm{b}}}{(m_{\mathrm{i}} + m_{\mathrm{b}})}$.

Typical relative collision velocities $\bar{v}$ and the corresponding relative collision energy $\bar{E} = 1/2 \, \mu \bar{v}^2 = 4 / \pi \times k_B T$ for fully thermalized impurities lie deep in the particle-like collision regime of the excitation spectrum. 
Therefore, when evaluating impurity-condensate collision rates, we do not expect significantly different behavior with respect to a fully classical, thermal bosonic bath.
A significant change of the collisional properties can be expected for energies $k_B T$ in the order of $1/2 \, \mu v_c^2$ \cite{Lausch2018}, corresponding to $k \approx 1 / \xi$, thus allowing to tune collisional properties by the choice of density, interaction and temperature of the BEC.
This is in contrast to studies of Bose polarons in a similar system, where only the low $k$ part of the impurities's spectral function is measured by RF spectroscopy, implying the scattering with low-momentum Bogoliubov excitations only. \cite{Jorgensen2016}. 
Thus, we use the effective elastic scattering length $a = 645\,a_0$ \cite{Takekoshi2012} ($a_0$ is the Bohr radius) to calculate the elastic collision rate $\Gammael = \sigma \bar{v} \nexp$ with the s-wave scattering cross section $\sigma = 4 \pi a^2$ for distinguishable particles.
	
A comparison of the rate constants of the elastic collision constant $\Gel$ with $\Gel = \sigma \bar{v} = \SI{1.61e-10}{\hertz \, \centi \meter^3}$ (for $T=\SI{300}{\nano \kelvin}$) and the spin-exchange constant $G = \nicefrac{\spinrate}{\left<n\right>}$ rates from our model yields information about the microscopic dynamics in the system.
The ratio $\nicefrac{\Gel}{G} \approx 11$ at the Rb bath temperature of $\SI{300}{\nano \kelvin}$ means that, on average, 1 in 11 collisions between impurity and bath atoms results in a spin-exchange. 
Since only few elastic collisions suffice for Cs atoms to thermalize in the Rb bath \cite{Davis1995, Hohmann2017}, fully thermalized Cs impurities in the BEC are assumed for modeling spin-exchange in the following.
The thermalization also implies that the density distribution of each impurity within one lattice well is effectively two-dimensional due to the large axial trap frequency in the lattice:
The energy level spacing in the lattice $\hbar \omega_{\mathrm{ax}} / k_B = \SI{3}{\micro \kelvin}$ is one order of magnitude above the BEC temperature of $\SI{300}{\nano \kelvin}$, therefore yielding negligible occupation of excited states in the lattice. \\

\textbf{Hamiltonian.} We consider a Cs (Rb) atom with total angular momentum $\mathbf{F}_{\mathrm{i}}$ ($\mathbf{F}_{\mathrm{b}}$). 
The quantum numbers are $\Fimp=3$ (hyperfine ground state) or $\Fimp=4$ for Cs and $\Fbath=1$ for Rb with the projections onto the quantization axis $\mFimp$ and $\mFbath$, respectively.
The full Hamiltonian of the interacting particles in the center-of-mass system writes \cite{Stoof1988}
\begin{displaymath}
 H = \frac{\mathbf{p}^2}{2 \mu} + \sum_{j=\mathrm{i}, \mathrm{b}} H^{\mathrm{i}}_{j} + \hint.
\end{displaymath}
Here, the first term is the total kinetic energy in the system (with relative momentum $\mathbf{p}$).
$H^{\mathrm{i}}_j = V_j^{\text{HFS}} + V_j^{\text{Z}}$ is the internal energy of each collision partner $j$ (impurity $\mathrm{i}$ and bath atom $\mathrm{b}$) with hyperfine and Zeeman energy  $V_j^{\text{HFS}}$ and $V_j^{\text{Z}}$, respectively.
Finally, $\hint$ is the interaction term that originates from a central interaction potential. Due to the central character of the interaction the total spin in the system $\mathbf{F} = \mathbf{F}_{\mathrm{i}} + \mathbf{F}_{\mathrm{b}}$ is conserved and $F$ and the projection to the quantization axis $M$ are good quantum numbers.
The impurity in state $\ket{\psi}_{\mathrm{i}}=\ket{\Fimp, \mFimp}$ and bath in state $\ket{\psi}_{\mathrm{b}}=\ket{\Fbath \mFbath}$ couple to $\ket{\Fimp \Fbath; F M}$ during the collision.
In order to calculate eigenstates of the Hamiltonian and the collision rates in the system, the interaction $\hint$ is expressed in terms of the total spin $F$, ranging from $\left| \Fimp - \Fbath \right|$ to $|\Fimp + \Fbath|$ as
\newcommand{\sumF}{\sum_{F=\left| \Fimp - \Fbath \right|}^{\Fimp + \Fbath}}
\begin{equation}
\hint = \nexp \sumF g_F \mathcal{P}_F
\label{eq:interactionHamiltonianProjection}
\end{equation}
with the projection operators onto total $F$, $\mathcal{P}_F = \sum_{M=-F}^{F} \ket{\Fimp \Fbath; F, M}\bra{\Fimp \Fbath; F, M}$ and the spatial wave function overlap $\nexp$. $g_F = ({4 \pi \hbar^2 }/{\mu}) \, a_F$ is the coupling constant with s-wave scattering length $a_F$ in scattering channels with total spin $F$.\\

\textbf{Elastic collisions.}
The central interaction potential leads to elastic collisions, where the atoms' internal state does not change.
However, the energy expectation values are different for each combination of internal states of the impurity $\ket{\psi}_{\mathrm{i}}$ and bath atom $\ket{\psi}_{\mathrm{b}}$.
For $B=0$ and $T=0$ the full Hamiltonian reduces to $\hint$, and the energy writes $E_{\ket{\psi}_{\mathrm{i}}} = \braket{\psi_{\mathrm{i}} \otimes \psi_{\mathrm{b}}|\hint|\psi_{\mathrm{i}} \otimes \psi_{\mathrm{b}} }$, with $\ket{\psi_{\mathrm{i}} \otimes \psi_{\mathrm{b}}} = \ket{{\Fimp \mFimp; \Fbath \mFbath}}$\\

When considering the situation, where an impurity atom is prepared in a quantum superposition of internal states $\ket{\psi}_{\mathrm{i}} = {\left(\ket{g} + i\ket{e}  \right)} / {\sqrt{2}}$, the state-depending interaction energy leads to a dephasing of the qubit. 
In the spin-echo measurement (see fig.~4 of main text) we use $\ket{g}=\ket{\Fimp=3, \mFimp=3}$ and $\ket{e}=\ket{\Fimp=4, \mFimp=3}$).
Dephasing leads to information loss, when the qubit is used as an information carrier, but could also be used to extract information about the bath for probing applications.
While for the former, low dephasing rates are desired, for the latter a strong interaction is favorable.
The state dependent energies $E_{\ket{g}}$ and $E_{\ket{e}}$ are calculated for Rb in $\ket{\psi}_{\mathrm{b}} = \ket{\Fbath=1, \mFbath=1}$, as used in the measurement.
We evaluate the Clebsch-Gordon coefficients $\Braket{\Fimp \Fbath; F, M | {\Fimp \mFimp \Fbath \mFbath}}$ in eq.~(\ref{eq:interactionHamiltonianProjection}) and get 
\begin{align}
E_{\ket{g}} &= \frac{4 \pi \hbar^2 n_{\mathrm{Rb}}}{\mu} a_{g, 4}  \\
E_{\ket{e}} &= \frac{4 \pi \hbar^2 n_{\mathrm{Rb}}}{\mu} \left( \frac{1}{5}a_{e, 4} + \frac{4}{5} a_{e,5} \right).
\end{align}
Therefore, the impurity qubit is dephasing at the rate $\delta_{\ket{e}-\ket{g}} = \frac{E_{\ket{e}}-E_{\ket{g}}}{h}$, which reads
\begin{equation}
	\delta_{\ket{e}-\ket{g}} = \frac{4 \pi \hbar^2 n_{\mathrm{Rb}}}{\mu h} \left(a_{g,4} -\frac{1}{5}a_{e, 4} - \frac{4}{5} a_{e,5} \right)
\end{equation}
with Plack's constant $h$.
Thus dephasing rates can be expressed in terms of scattering lengths differences, here $\Delta a_{g-e} = a_{g,4} -\frac{1}{5}a_{e, 4} - \frac{4}{5} a_{e,5}$.
The state dependent scattering lengths are $a_{g, 4}=648\,a_0$ for the ground state and $a_{e, 4} = 570\,a_0$ and $a_{e, 5} =  626\,a_0$ in the excited state \cite{Tiemann2018}, yielding an effective scattering lengths difference for our choice of qubit states of $\Delta a_{g-e} = 33\,a_0$.
Analogously, the dephasing rates for all possible state combinations can be evaluated.
For example, for an alternative qubit choice $\ket{g}=\ket{\Fimp=3, \mFimp=0}$ and $\ket{e}=\ket{\Fimp=4, \mFimp=0}$, this yields $\Delta a_{0-0}= 330 \, a_0$.
\\

\textbf{Spin-exchange collisions.}
Additionally to elastic collisions, the central interaction potential also allows an exchange of angular momentum between the collision partners, i.e. spin-exchange, while maintaining the total projection $M=\mFimp + \mFbath$.

At ultracold temperatures and finite magnetic fields the spin-exchange is unidirectional, determined by the eigenenergies (Zeeman energy) of impurity and bath atoms.
For the Cs-Rb combination, in each spin-exchange collision the energy $\nicefrac{\Eexo}{B} = h \cdot \SI{350}{\kilo \Hz \per G}$ is converted into kinetic energy, while transferring $\SI{1}{\hbar}$ from the impurity to a bath atom.
The energy of $\Eexo = k_B \times \SI{4}{\micro \kelvin}$ (for $B=\SI{250}{\milli G}$) and angular momentum due to spin-exchange is transferred to a bath, consisting of $>10^4$ atoms.
For our strongly imbalanced mixture, this does effectively not change the temperature or the mean spin projection of the bath.\\

A calculation of the rate constant $G^{\ket{f}}_{\ket{i}}$ for a transition from state $\ket{i}=\ket{\Fimp \Fbath; \mFimp \mFbath}$ to state $\ket{f}=\ket{\Fimp \Fbath; \mFimp^{\prime} \mFbath^{\prime}}$ requires a full diagonalization of the system's Hamiltonian including kinetic, Zeeman and hyperfine energies. In general it yields a dependency on the initial $m_F$ states of both collisional partners \cite{Stoof1988}.

For the case of Rb in $\mFbath=-1$ being discussed here, the rates of spin-exchange have been calculated theoretically \cite{Tiemann2018}.
We expect a spin-exchange constant for an exchange of $1 \hbar$ between Cs and Rb of $\bar{G}_{1\hbar}=\SI{1.57(65)E-11}{\hertz \, \centi \meter^3}$ for the allowed transitions at low magnetic background fields $<\SI{1}{G}$ (for comparison to measurement, see sec. \ref{sec:rateConstant}).
Here, the error bar gives the standard deviation for $G$ values in different $\mFimp$ states.
Note, that spin-exchange is also allowed in quanta of $2\hbar$. Here, however, the rate constant $\bar{G}_{2\hbar}=\SI{1.22(76)E-12}{\hertz \, \centi \meter^3}$ is expected to be one order of magnitude lower than for spin-exchange of $1 \hbar$, so we neglect this process in our model.
\\

\subsection{Rb and Cs preparation}
\begin{figure}
	\centering
	\includegraphics[width=88mm]{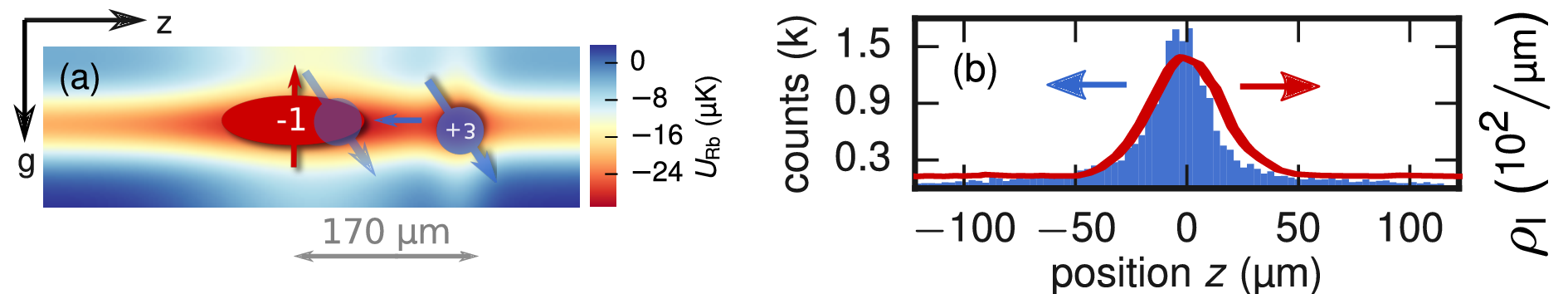}
	\caption{(a) Potential landscape of the Rb dipole trap $U_{\mathrm{Rb}} / k_B$.
		Cs (blue) is prepared independently from the BEC (red) and transported to the BEC subsequently.
		Numbers give the respective $m_F$ states.
		$z$ is the main experiment axis, and $g$ gives the direction of gravity.
		(b) Position distribution of Cs within the BEC after transport (blue histogram) and the measured BEC line density $\rho_{\mathrm{l}}$ at the beginning of the interaction time after $\SI{6.5}{\milli \second}$ expansion time.
		}
	\label{fig:S1}
\end{figure}
The Rb BEC is created  in an all-optical evaporation scheme in a crossed dipole trap at $\SI{1064}{\nano \meter}$ in the magnetic field insensitive $\mFbath=0$ state.
The atom number and condensate fraction is measured after a free expansion of $\SI{20}{\milli \second}$ in time-of-flight.

After evaporation, we increase the Rb trapping potential by increasing the power of the axial dipole trap adiabatically from $U_{0, \mathrm{Rb}} / k_B = \SI{-3}{\micro \kelvin}$ after evaporation to $U_{\mathrm{Rb}} / k_B = \SI{-30}{\micro \kelvin}$ (with Boltzmann constant $k_B$).
The BEC is characterized after this adiabatic compression, yielding a total atom number of $(10-20) \times 10^3$ and a condensate fraction of about 0.30 - 0.35.

Subsequently, Cs atoms are loaded from a high-gradient magneto-optical trap into an independent crossed dipole trap, sharing the main, horizontal dipole trap beam along the $z$ axis with the BEC (see fig. \ref{fig:S1}, (a)).
Three Raman sideband cooling pulses \cite{Treutlein2001}, separated by short evolution times of $\pi/2 \omega_\mathrm{rad}$ ($\omega_\mathrm{rad} = 2 \pi \times 600\,$Hz is the radial trapping frequency) cool the Cs atoms to approximately $\SI{2}{\micro \kelvin}$.
Rb atoms are transferred into one of the $\mFbath = 0, \pm 1$ states by two Landau-Zener microwave sweeps, near-resonant to the $h \times \SI{6.8}{\giga \hertz}$ hyperfine transition.
Figure \ref{fig:S1} shows the potential landscape for Rb together with the Rb and Cs position distributions along the main experiment axis $z$ after preparation.\\

\textbf{The dipole trap.}
The trap frequencies of the dipole trap in radial and axial direction for Rb are $\omega_r = 2 \pi \times \SI{700}{\hertz} $ and $\omega_z=2 \pi \times \SI{50}{\hertz}$, respectively.
Note, that Cs has nearly the same trapping frequencies due to the favorable ratio of mass and dipole force, leading to a negligible gravitational sag between both species below $\SI{50}{\nano \meter}$.
The initial density of a BEC with condensate fraction $\eta$, temperature $T$ and total atom number $N$ in the trap is calculated as a bimodal distribution $n_{\mathrm{b}} = n_{\mathrm{th}}(r, z) + n_{\mathrm{TF}}(r, z)$. Here, $n_{\mathrm{th}}(r, z)$ is the density of the thermal background and $n_{\mathrm{TF}}(r, z)$ the Thomas-Fermi distributed condensed fraction  \cite{Dalfovo1999}.
The thermal density profile writes
\newcommand{\denspeakthermal}{n_{0, \mathrm{th}} }
\begin{equation}
	n_{\mathrm{th}}(r, z) = \denspeakthermal \exp{\left(  
		-\frac{r^2}{2 \sigma_r^2} - \frac{z^2}{2 \sigma_z^2}
		\right)}
\end{equation}
with the thermal peak density $\denspeakthermal = \frac{(1-\eta)N}{(2\pi)^{\nicefrac{3}{2}} \sigma_r^2 \sigma_z}$ and widths of the thermal cloud in radial and axial direction $\sigma_r = \sqrt{\frac{k_B T }{m_{\mathrm{b}}}} \frac{1}{\omega_r}$ and $\sigma_z = \sqrt{\frac{k_B T }{m_{\mathrm{b}}}} \frac{1}{\omega_z}$.
The Thomas-Fermi density profile $n_{\mathrm{TF}}(\mathbf{r})$ of a BEC writes 
\newcommand{\denspeakBEC}{n_{0, \mathrm{TF}} }
\begin{equation}
	n_{\mathrm{TF}}(r, z) = \denspeakBEC \left( 1 - \frac{r^2}{R_r^2} - \frac{z^2}{R_z^2} \right)
\end{equation}
with peak density $\denspeakBEC = \frac{15 \eta N}{8 \pi R_r^2 R_z}$ and Thomas-Fermi radii $R_{r, z} = \frac{2\mu_c}{m \omega_{r, z}^2}$ in radial and axial direction, respectively. Here, $\mu_c = \frac{15^\frac{2}{5}}{2} \left( \frac{\eta N a}{\bar a} \right)^\frac{2}{5} \hbar \bar\omega$ is the chemical potential of an interacting BEC with scattering length $a = 101\,a_0$ \cite{Kempen2002} and characteristic length $\bar a = \sqrt{\frac{\hbar}{m_{\mathrm{b}} \bar \omega }}$, $ \bar{\omega} = (\omega_r^2 \omega_z)^{1/3}$. 
Typically, our BEC has a calculated peak density on the order of $\SI{1e14}{\centi \meter^{-3}}$ and Thomas-Fermi radii of $R_r=\SI{1}{\micro \meter}$ and $R_z=\SI{10}{\micro \meter}$.\\

\textbf{Loss channels of Cs.}
Depending on the Cs hyperfine state $\Fimp=3,4$, different loss channels limit the lifetime of Cs atoms in the Rb BEC.
For Cs in the $\Fimp=3$ hyperfine ground state, three-body recombination of one Cs atom with two Rb atoms leads to a loss at a rate $\Lambda_{\mathrm{3body}} = L_3 \left< n^2 \right>$ with $\left<n^2\right> = \int n_{\mathrm{b}}^2 n_{\mathrm{i}} \mathrm{d}^3 \mathbf{r}$.
The value of $L_3 = \SI{28(1)e-26}{\hertz \, \centi \meter^6 }$ has been experimentally obtained for Rb in $\mFbath=0$ in an independent measurement. 

When Cs is prepared in the excited hyperfine state $\Fbath=4$, additional 2-body recombination can occur at a rate $\Lambda_{\mathrm{2body}} = L_2 \left< n\right>$ with a loss coefficient $L_2 = \SI{4(2)e-12}{\centi \meter^3 \hertz}$ (determined for Rb in $\mFbath=0$) and the (linear) density overlap $\left< n \right>$.

Since the expected rate of three-body loss $\Lambda_{\mathrm{3body}}=\SI{16}{\kilo \hertz}$ is in the same order of magnitude as the elastic collision rate $\Gammael = \SI{36}{\kilo \hertz}$, we intend to reduce the loss rate in order to observe Cs-BEC dynamics driven by elastic and spin-exchange dynamics, rather than mere loss of Cs.\\

Therefore, before initiating the interaction of Cs and Rb, the axial confinement is lowered by switching off the axial dipole trap beam, so the axial trap frequency $\omega_z$ reduces to $\tilde{\omega}_z = 2\pi \times \SI{8}{\hertz}$ instantaneously.
The Thomas-Fermi radii evolve according to $R_{r, z}(t) = \lambda_{r,z}(t) R_{r, z}(0)$ with time-dependent proportionality factors $\lambda_{r,z}$(t) \cite{Kagan1996, Castin1996}, given by
\begin{equation}
	\ddot{\lambda}_{j=r,z} = \frac{\omega_{j}(0)^2}{\lambda_{j} \lambda_r^2 \lambda_z} - \omega_{j}^2(t) \lambda_{j},
\end{equation}
with $\lambda_{r,z}(0)=1$ and $\dot{\lambda}_{r,z}(0)=0$ and the time dependent trap frequencies $\omega_{r,z}(t)$.
During the expansion, the Thomas-Fermi radius $R_z$ increases from initially $\SI{13}{\micro \meter}$ to $\SI{270}{\micro \meter}$ axially, while the peak density reduces by almost a factor of ten within the total BEC expansion of up to $\SI{26.5}{\milli \second}$.

For the thermal background, the degrees of freedom do not couple in the quasi-harmonic trapping potential, so the radial position distribution remains unaffected.
Due to the short expansion time $<\SI{26}{\milli \second}$ with respect to the trap period of $\SI{125}{\milli \second}$, we assume free expansion along the $z$ axis with \cite{ketterle1999}
\begin{equation}
	\sigma_z^2(t) = \sigma_z^2(0) + \frac{k_B T}{m_{\mathrm{b}}} t^2.
\end{equation}
We compare our model to a measurement of the line density $\rho_{\mathrm{l}}$ of the Rb bath (BEC and thermal background) and find good agreement until interaction times of about $\SI{15}{\milli \second}$. 
For longer interaction times, $\rho_{\mathrm{l}}$ shows a bimodal density distribution, which we cannot reproduce by our model.
We attribute the occurrence of this localized fraction to the emergence of a shallow lattice in the axial dipole trap beam due to an unwanted, partial retro-reflection of the trapping light on the glass cell, which might induce a localization in the lowest Bloch band of that lattice \cite{Kolovsky2004}.
We estimate a lattice depth on the order of $0.2 \, E_r$ (recoil energy $E_r=\nicefrac{(\hbar k)^2}{2 m_{\mathrm{b}}}$ with wave vector $k=\nicefrac{2 \pi}{\SI{1064}{\nano \meter}}$).
Since lattice effects are only expected to occur along the $z$ direction, we expect the radial distribution to remain unaffected.\\
\begin{figure}
	\centering
	\includegraphics[width=85mm]{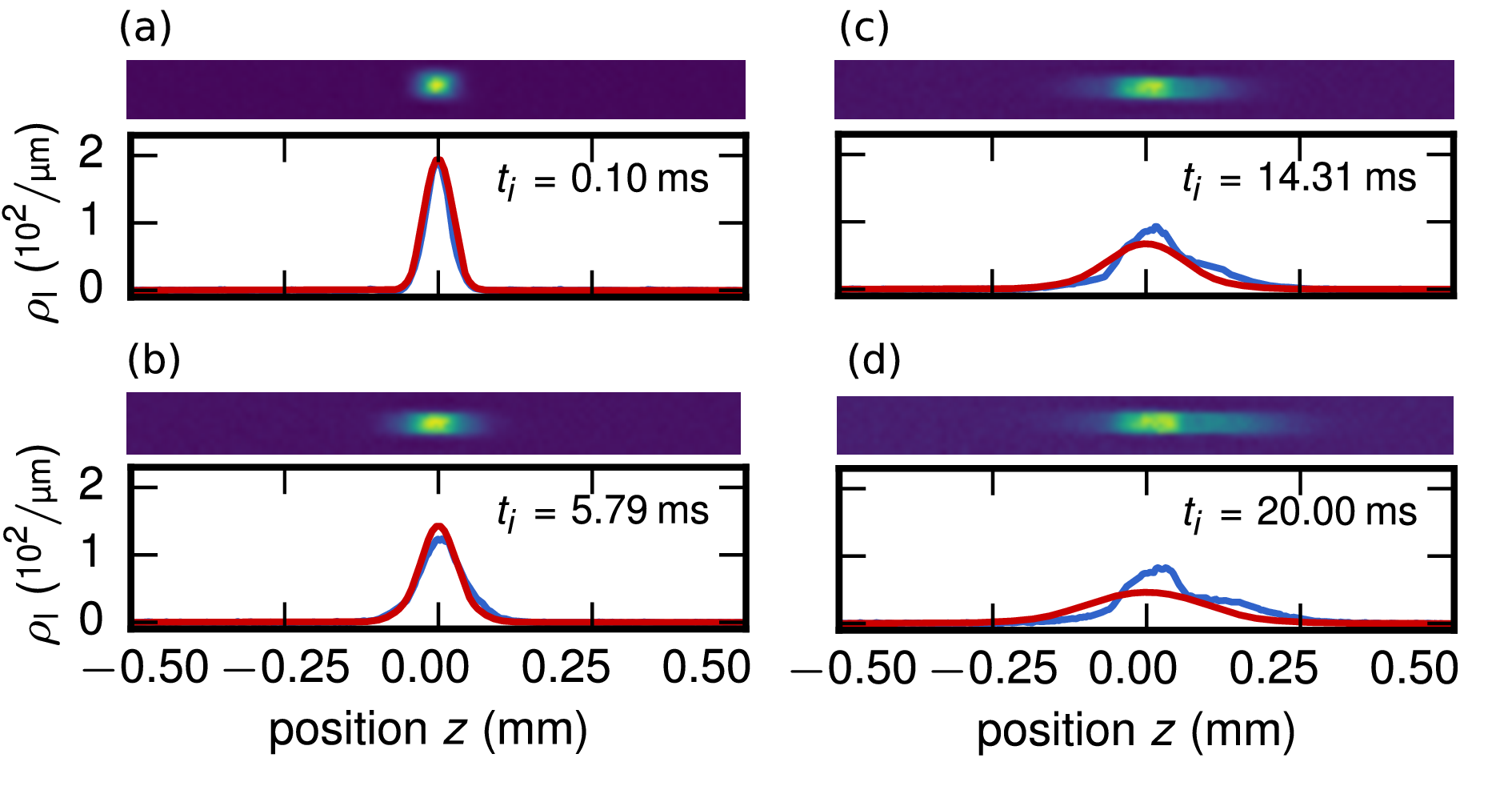}
	\caption{Quasi \textit{in situ} atomic BEC density and extracted line density $\rho_{\mathrm{l}}$ of the BEC after $\SI{1.5}{\milli \second}$ time-of-flight (atom number 14$\times 10^3$ atoms and condensate fraction 0.35).
		(a) - (d) show atomic density and extracted line densities (blue) for various interaction times $t_i$ given in the respective panel together with our model (red).
		The line density is extracted by vertically binning the measured atomic column density.
		The model takes the limited imaging resolution of the absorption imaging system of $\SI{10}{\micro \meter}$ into account.}
	\label{fig:S2}
\end{figure}

\textbf{Cs distribution.}
Cs atoms are pinned to their position by the species-selective lattice, yielding trap frequencies of $\omega_r=2\pi \times \SI{715}{\hertz}$ and $\omega_z = 2 \pi \times \SI{63}{\kilo \hertz}$.
Since the mean free path length $\nicefrac{1}{( n_{\mathrm{b}}(t_i) \sigma )}$ of Cs impurities in the expanding BEC exceeds the radial size of the BEC for all interaction times $t_i$, we expect no localization effects in the Rb bath in radial direction. 
At a bath temperature of $\SI{300}{\nano \kelvin}$, the Cs distribution has a spatial extent of $\sigma_r=\SI{1}{\micro \meter}$ in radial direction.
Along the axial direction ($z$) impurities predominantly occupy the ground state of the species-selective lattice with a width of $\sigma_z \approx \SI{10}{\nano \meter}$.

\subsection{Species-selective lattice potential}
The species-selective lattice is formed by two counter-propagating, linearly polarized laser beams at a wavelength of $\lambda_{\mathrm{Lat}} = \SI{790}{\nano \meter}$, superposed to the axial dipole trap along $z$.
The wavelength choice realizes a tune-out trapping scheme, exploiting the coupling to both Rb-$D$ lines \cite{LeBlanc2007}.
A selectivity of $1800$ is achieved for Rb in the $\mFbath=\pm1$ state, limited by vector light shifts \cite{Schmidt2016_2}.
A small detuning of the laser frequency can be introduced, allowing the transport of Cs atoms in this conveyor belt lattice \cite{Kuhr2001}.
For the transport, we use a lattice potential of $\SI{150}{E_{r, \mathrm{i}}}$ for the impurity atoms with a residual potential of $\SI{0.05}{E_{r, \mathrm{b}}}$ for the bath atoms  (with photon recoil energy $E_{r, \mathrm{i, b}}$ for impurity and bath atoms, respectively).\\

During transport ($\SI{10}{\milli \second}$ duration) and holding of Cs impurities in the species-selective lattice ($\SI{20}{\milli \second}$ duration) the lattice causes an off-resonant photon scattering of on average $0.25$ by each BEC atom.
All BEC characteristics given in respective measurements are determined including this off-resonant photon scattering.

\subsection{Impurity Spin Readout}
\begin{figure}
	\centering
	\includegraphics[width=75mm]{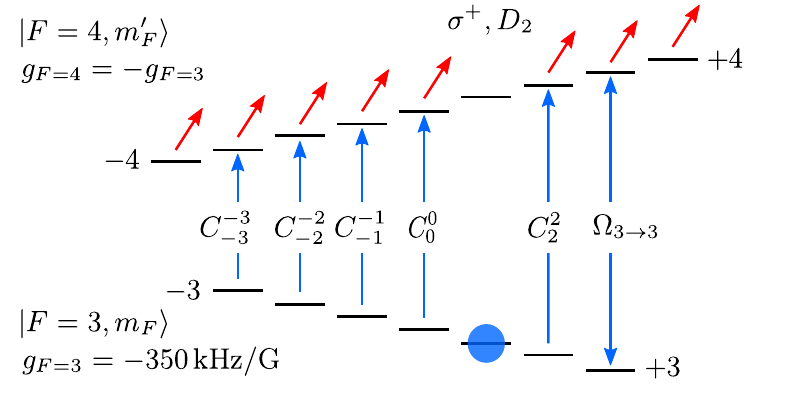}
	\caption{Detection scheme of $\mFimp$ population. 
			The population of one $\ket{\Fimp=3, \tilde{m}_{F, \mathrm{i}}}$ state (here $\tilde{m}_{F, \mathrm{i}}=1$) is measured by transferring the population of all other $\ket{\Fimp=3, \mFimp}$ states in 6 independent Landau-Zener (LZ) microwave transitions to the excited hyperfine state. 
		For the LZ sweeps $\pi$ transitions ($\Delta \mFimp = 0$) are used.
		For details, see text.
	}
	\label{fig:Landau-Zener}
\end{figure}

While the $\mFimp$ population in ultracold gases is routinely detected in Stern-Gerlach experiments during time-of-flight, we rely on \textit{in situ} fluorescence imaging of Cs atoms in the dipole trap, which excludes those standard methods.
In contrast, our $\mFimp$ mapping scheme is based on microwave transitions between the hyperfine ground states $F=3$ and $F=4$ of Cs (see fig.\ref{fig:Landau-Zener}), while Cs atoms remain localized in the species-selective lattice.

The population of a desired $\tilde{m}_{F, \mathrm{i}}$ state is measured in two steps.
First, the $\mFimp \neq \tilde{m}_{F, \mathrm{i}}$ states are transferred to $\ket{\Fimp=4, \mFimp^{\prime}}$ by independent Landau-Zener (LZ) microwave sweeps, near-resonant to the Cs $\SI{9.2}{\giga \hertz}$ clock transition.
In order to guarantee adiabatic transfer for all $\mFimp$ states, the Rabi frequency of the transition $\ket{\Fimp=3, \mFimp=3} \rightarrow \ket{F=4, \mFimp^{\prime}=3}$ has been measured.
The remaining Rabi frequencies $\Omega_{\mFimp \rightarrow \mFimp^{\prime}}$ were calculated based on the ratio of their transition strength $C_{\mFimp}^{\mFimp^{\prime}}$ to the one of the $\Omega_{\mFimp=3\rightarrow \mFimp^{\prime}=3}$ transition (see fig. \ref{fig:Landau-Zener}).
After the LZ transitions have been completed, the population in the $\ket{F=4}$ manifold is removed by a state selective push-out light pulse on the $D_2, F=4 \rightarrow F=5$ cycling transition,
leaving only $\tilde{m}_{F, \mathrm{i}}$ atoms in the trap.

\subsection{Spin-Evolution model}
We model the evolution of the impurity spin with a rate equation, where spin-exchange at rate $\spinrate$ and atom loss due to inelastic three body loss $\Lambda$ change the population in each $\mFimp$ state $N_{\mFimp}$ according to
\begin{equation}
\begin{aligned}
\dot{N}_{+3} &= -\left( \spinrate + \lossrate \right) N_{+3} \\
\dot{N}_{ \mFimp } &= -\left( \spinrate + \lossrate \right) N_{\mFimp} + \spinrate N_{\mFimp+1} \\ & \quad \text{for} \quad \mFimp = -2, -1, 0, +1, +2 \\
\dot{N}_{-3} &=  \spinrate N_{-2} - \lossrate N_{-3}.
\end{aligned}
\label{eq:rateModel}
\end{equation}
\textbf{Average Rates}
A common approach to solve the spin dynamics uses an average spin-exchange rate $\left<\spinrate\right>$.
Here, the time dependent density overlap $\nexp(t) = \int n_{\mathrm{i}}(\mathbf{r}) n_{\mathrm{b}}(\mathbf{r}, t) \mathrm{d}^3\bf{r}$ of impurity atoms and the expanding BEC is calculated. 
This approach is used, when expectation values for spin-exchange rates in the main body, as well as in this appendix are given.\\

\textbf{Monte-Carlo approach}
The use of averaged rates $\left< \spinrate \right>$ however neglects the influence of the inhomogeneous density distribution of impurities within the Rb bath and of the bath itself, which both lead to a temporal fluctuation of the spin-exchange rate for the model, leading to an effective broadening of the $\mFimp$ distribution for increasing interaction durations $t_i$.
In order to include the influence of the inhomogeneous distributions, we use a Monte-Carlo simulation, where the local density of the Rb cloud $n_{\mathrm{b}}(\mathbf{r}, t)$ is evaluated for each Cs impurity and time step in the Monte-Carlo sample individually.
The rate model (see eq.~(\ref{eq:rateModel})) for a sample of $N_{\mathrm{tot}}$ independent realizations (impurity atoms) is solved, where in each integration step at time $t$ the position $\mathbf{r}_0(t)$ of the impurity atom $j$ is randomly drawn from its thermal distribution in the trap.
This yields an impurity density $\nimp(\mathbf{r})=\delta(\mathbf{r} - \mathbf{r}_0)$ and therefore density overlap with the BEC of $\nexp = \nbath(\mathbf{r}_0)$ for each time step $t$, so the spin-exchange rate writes
\begin{equation}
{\spinrate}(t)	= G \, \nbath(\mathbf{r}_{0}(t)).
\label{eq:localGamma}
\end{equation}
By independently solving the rate equation for each of all $N_{\mathrm{tot}}$ impurities, this yields the $\mFimp$ population $N^{(j)}_{\mFimp}(t)$, from which the population in the ensemble is calculated as
\begin{equation}
N^{}_{\mFimp}(t) = \sum_{j=1}^{N_{\mathrm{tot}}} N^{(j)}_{\mFimp}(t)
\label{eq:spinEvolutionTotal}
\end{equation}
in each time step.
When solving the rate model, initial $\mFimp$ populations obtained from the respective measurement.
The Monte-Carlo model is used for all analyses of spin-evolution in the text body (Fig.~2, Fig.~3) as well as this appendix (Fig.~S5).

\subsection{Measuring spin-exchange constant $G$}
\label{sec:rateConstant}

\begin{figure}
	\centering
	\includegraphics[width=75mm]{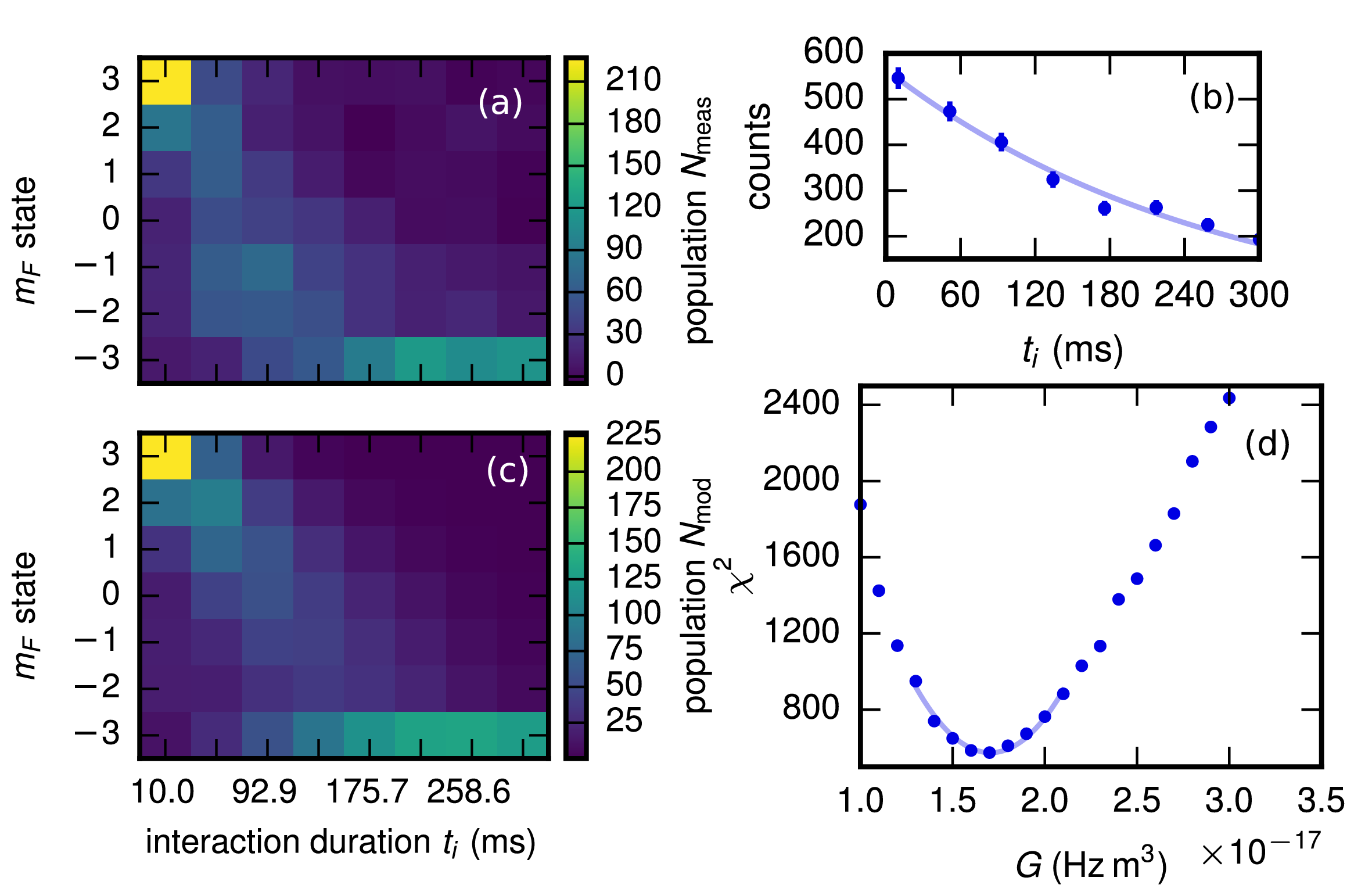}
	\caption{Extracting the spin-exchange coefficient $G$. 
		(a) Measurement of the spin evolution of Cs atoms prepared in a thermal Rb bath (15(1)\,$\times 10^3$ atoms, temperature $T=\SI{1.25(10)}{\micro \kelvin}$) at $B=\SI{750}{\milli G}$. 
		The population $N_{\mathrm{meas}}$ gives the total number of detected Cs atoms for one specific experimental setting ($t_i,\, m_F\,\mathrm{state}$). The data was taken throughout 4037 independent experimental runs with a duration of about $\SI{10}{\second}$ each.
		(b) The total population decays due to three-body recombination during the interaction time $t_i$.
		Here, total population (\enquote{counts}) refers to the sum of all $\mFimp$ populations for the respective interaction duration $t_i$.
		From the population decay, the loss rate $\Lambda_{\mathrm{3body}} = \nicefrac{1}{\SI{265}{\milli \second}}$ for the spin-exchange model (eq. \ref{eq:rateModel}) is determined.
		(c) Modeled $\mFimp$ evolution from the Monte-Carlo simulation using the initial $\mFimp$ population from the measurement (a) as the starting distribution.
		Here, the spin-exchange constant $G=\SI{1.71e-11}{\hertz \, \centi \meter^3}$ is used, which fits best with the measurement (see d) 
		(d) A $\chi^2$ optimization is used to extract the rate constant $G$.
	}
	\label{fig:supp3}
\end{figure}

We apply the Monte-Carlo model to determine the spin-exchange constant $G$ in our atomic mixture. 
Therefore, a dilute cloud of Rb atoms is prepared in the $\mFbath=-1$ state, so spin-exchange can be observed in a classical bath with well known density distribution.
We combine Cs and Rb as described in \cite{Hohmann2017} and allow Cs to fully thermalize within the dilute Rb cloud.
We measure the $\mFimp$ dynamics for $\SI{300}{\milli \second}$ at a magnetic background field of $B = \SI{750}{\milli G}$ (see fig.\ref{fig:supp3}) and observe spin-exchange with the bath, as well as atom loss, presumably due to three-body recombination.
Models for different $G$ constants are compared.
For each data point (pixel) in the measurement, a chi-squared value is calculated $\chi^2_{j}=\frac{(N_{j, \mathrm{exp}} - N_{j, \mathrm{mod}})}{\sigma_j^2}$ from the measured population $N_{j, \mathrm{exp}}$, the modeled population $N_{j, \mathrm{mod}}$ and the expected uncertainty $\sigma_j$.
We extract a spin-exchange constant $G = \SI{1.71e-11}{\hertz \, \centi \meter^3}$ by minimizing the total $\chi^2 = \sum_j \chi^2_j$.
The minimization yields a statistical uncertainty of $\Delta G_{\mathrm{stat}} = \SI{0.01e-11}{\hertz \, \centi \meter^3}$.
We compare our result to the theoretically estimated spin-exchange constant $\SI{1.57E-11}{\hertz \, \centi \meter^{3}}$ with an uncertainty of $\Delta G = \SI{0.61e-11}{\hertz \, \centi \meter^3}$ originating from the $\mFimp$ dependency (see sec. \ref{sec:interactionPotential}) and we find good agreement. 
Discrepancies between the measurement and our model occur mainly for short interaction durations, which we attribute to our assumption of $\mFimp$-independent spin-exchange.
In fact, the theoretical calculation of $\SI{1.57(61)E-11}{\hertz \, \centi \meter^{3}}$ shows a larger uncertainty $\Delta G$ due to the state dependency than our $\chi^2$ fit $\Delta G_{\mathrm{stat}}$.
Therefore, when referring to $G$, we use the uncertainty of the theoretical value.
In our mixture with Rb in $\mFbath=-1$, we do not expect a strong dependency of the spin-exchange constant on the magnetic field \cite{Tiemann2018}, e.g. due to Feshbach resonances, so we use the extracted $G$ constant for modeling spin-evolution at $B=\SI{250}{\milli G}$.

\subsection{Details on spin-echo measurement}
\begin{figure}
	\centering
	\includegraphics[width=75mm]{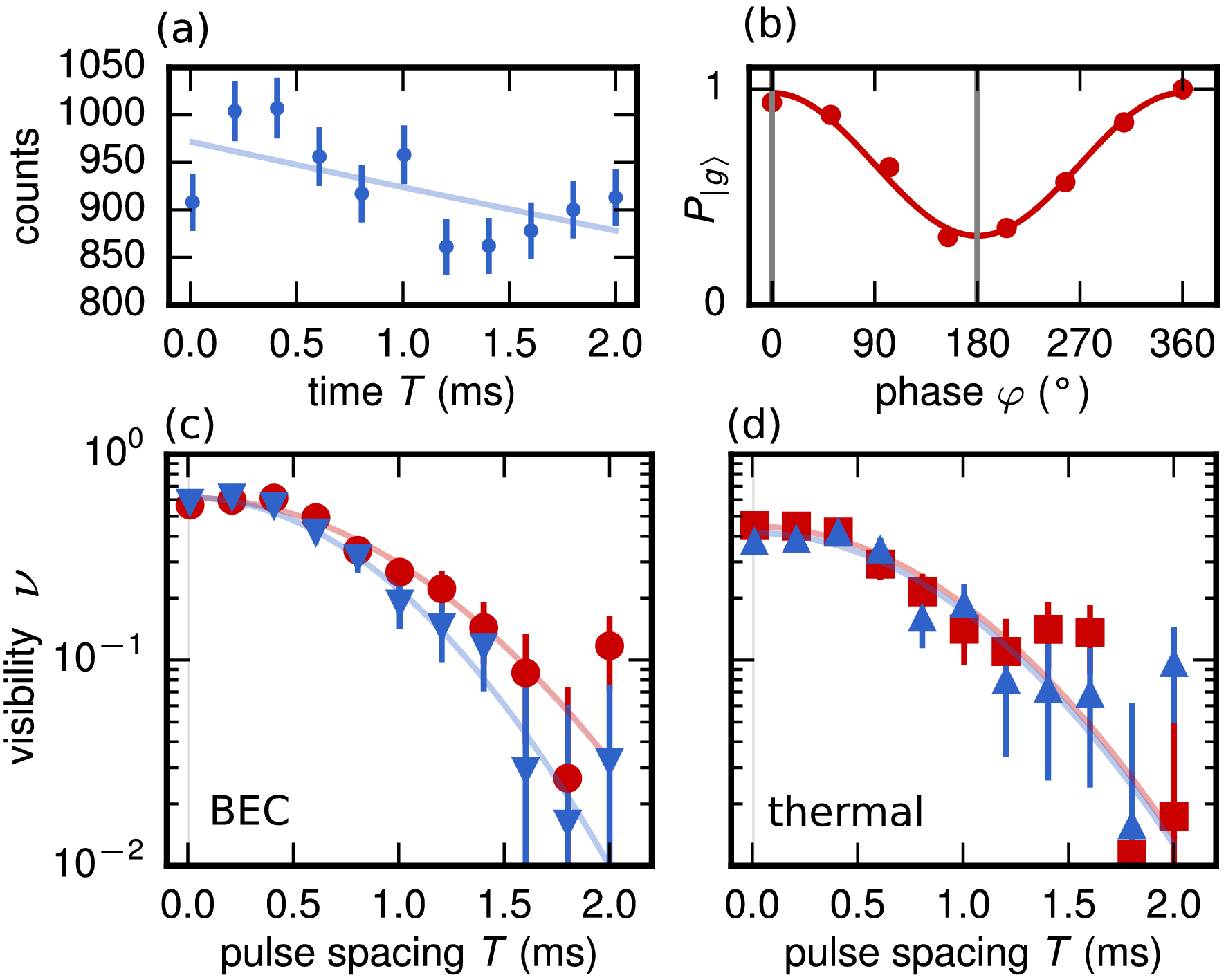}
	\caption{Measurement of the spin-echo contrast.
		(a) The total number of Cs atoms for each detected interaction duration for highest density (BEC) is shown.
		(b) Spin-echo fringe for a constant interaction time (here $T=\SI{100}{\micro \second}$) and varied pulse phase $\varphi$ of the last $\nicefrac{\pi}{2}$-pulse.
		The relative population $P_{\ket{g}}$ of the ground state $\ket{g}$ is determined by removing atoms in $\ket{e}$ from the trap with a resonant light pulse.
		For measurements in (c) and (d) the visibility $\nu$ is extracted from populations $P_{\ket{g}}$ at phases $\varphi=0, \SI{180}{\degree}$ (see vertical lines) as  $\nu = {|P_{\varphi=\SI{180}{\degree}}-P_{\varphi=\SI{0}{\degree}}|} / 
		{(P_{\varphi=\SI{180}{\degree}}+P_{\varphi=\SI{0}{\degree}})}$.
		(c) For impurities prepared in a BEC (red circles), we measure a coherence time of $T_2=\SI{1.17 \pm .06}{\milli \second}$.
		In the case, when Cs is prepared in the BEC, but the BEC is removed before the coherence measurement (blue triangles), we get $T_2=\SI{0.98\pm 0.03}{\milli \second}$.
		(d) Cs impurities are prepared in a thermal Rb bath (spin mixture) slightly above the critical temperature.
		We measure the same coherence time, when Rb is present (red squares, $T_2=\SI{1.07\pm 0.08}{\milli \second}$) and when Rb is removed before the pulse-echo sequence (blue triangles, $T_2=\SI{1.07\pm 0.10}{\milli \second}$).
		The initial visibility is limited due to imperfections in the initial $\mFimp$ preparation and spin-exchange before the start of the spin-echo measurement (in the thermal case), both yielding a measurement background.
		Error bars give statistical uncertainties in the atom number determination.
	}
	\label{fig:S4}
\end{figure}
We study the coherence properties of Cs impurities immersed into the BEC in a spin-echo sequence.
The coherence of individual Cs atoms in a similar system, but without a bath, has been studied in the work of Kuhr \textit{et al.} \cite{Kuhr2005}.\\
Experimentally, Cs atoms are immersed into the BEC $\SI{6}{\milli \second}$ before the spin-echo sequence in order to ensure a thermalization of Cs atoms within the BEC.
We create a quantum superposition $\ket{\psi}_{\mathrm{i}}  = \nicefrac{1}{\sqrt{2}}(\ket{g}+i\ket{e})$ with $\ket{g} = \ket{\Fimp=3, \mFimp=3}$ and $\ket{e} = \ket{\Fimp=4, \mFimp^{\prime}=3}$.
The two states have been chosen due to the strong resonant Rabi coupling frequency of $\Omega_0 = 2 \pi \times\SI{40}{\kilo \hertz}$ in our setup.
During the measurement, we apply a magnetic background field of $\SI{250}{\milli G}$ along the $z$ axis.
The coherence time is extracted by measuring the ground state population after the last $\pi / 2$ pulse, when varying the phase of the last pulse. From the visibility decay (for details, fig.~\ref{fig:S4}), then the coherence time is extracted. 
First, we probe the coherence of individual impurities in a purely thermal Rb gas at a temperature of approximately $300\,$nK.
When Rb is removed from the trap just before the spin-echo sequence, i.e. in the absence of impurity-bath interactions, we measure a coherence time of $T_2=\SI{1.07\pm 0.10}{\milli \second}$.
By contrast, when Rb was present, we obtain $T_2=\SI{1.07\pm 0.08}{\milli \second}$.
We compare the coherence time in a thermal bath to a situation, where the Cs impurities are transported into the Rb BEC of comparable temperature, but much higher density.
The BEC is prepared in the $\mFbath=1$ state, preventing an influence of spin-exchange collisions on the coherence.
Here, we extract a coherence time of $\SI{1.17\pm0.06}{\milli \second}$, when Rb is present during the pulse-echo sequence.
This means that coherence is maintained despite elastic impurity-bath collisions at a rate of  $\nicefrac{1}{\SI{140}{\micro \second}}$ for the highest density (BEC) during the spin-echo sequence.
Finally, when the BEC is removed from the trap before the pulse-echo sequence, the coherence time slightly reduces to $T_2=\SI{0.98\pm0.03}{\milli \second}$, which we attribute to a heating effect from the push-out process:
During the push-out, Rb is accelerated by a resonant light beam in radial direction. 
After a push-out duration of $\SI{20}{\micro \second}$, we expect Cs and Rb to be fully separated ($\SI{3}{\micro \meter}$ distance), while Rb is acquiring a kinetic energy of $\SI{11}{\micro \kelvin}$.
The acceleration of Rb enhances the collisional cross section, and at a density overlap of  $\nexp = \SI{2.7e13}{\centi \meter^{-3}}$, one in three Cs atoms undergoes a collision with Rb, leading to heating and thereby dephasing fluctuations. \\

In the following, we discuss different decoherence mechanisms, while the results are further discussed in the body of this work.
In our experiment, we do not expect longitudinal decay ($T_1$) on relevant time scales, since the transition $\Fimp=4 \rightarrow \Fimp=3$ is dipole-forbidden.
However, longitudinal decoherence might be mimicked by two-body (hyperfine-changing) relaxation of the $\Fimp=4$ state into the ground state $F=3$, leading to atom loss and reduced contrast.

For the calculated Cs-Rb overlap during the pulse-echo sequence of $\nexp = \SI{2.7e13}{\centi \meter^{-3}}$ two-body loss is expected to yield a lifetime of the $F=4$ state of $\tau_{\mathrm{2body}} > \SI{10\pm5}{\milli \second}$ in agreement with our observation (see fig. \ref{fig:S4}~(b)), which is long compared to the extracted coherence time in the measurement. \\

The transverse coherence time $T_2$ of the atomic ensemble is limited by inhomogeneous, but quasi-constant dephasing ($T_2^{*}$) and a fluctuation of the dephasing ($\Ttwoprime$) as $\nicefrac{1}{T_2} = \nicefrac{1}{\Ttwoprime}+\nicefrac{1}{T_2^{*}}$ (see e.g. \cite{Kuhr2005}).
In our measurement, we use a spin-echo technique in order to find the fundamental limitation of the coherence.
Therefore, the measured $T_2$ directly yields $\Ttwoprime$.
$\Ttwoprime$ is estimated by the influence of temporal fluctuation $\Delta \delta = \sqrt{2} / \Ttwoprime$ of the detuning $\delta(t) = \omega_0(t) - \omega_L(t)$ (with atomic transition frequency $\omega_0$ and microwave driving frequency $\omega_L$).
In our system, we expect magnetic field fluctuations to be main contributors to $\Delta \delta$. 
For our combination of Zeeman states of $\ket{g}$ and $\ket{e}$, a fluctuation of $\Delta \delta = \SI{1}{\kilo \hertz}$ is induced by magnetic field fluctuations of $\Delta B = \SI{0.48}{\milli G}$, which equals roughly $\SI{0.1}{\percent}$ of the earth magnetic field in our laboratory.
Additional dephasing sources are fluctuations connected to the finite temperature of the atoms, heating from to dipole traps \cite{Kuhr2005}, as well as atomic collisions.\\

\end{document}